# The Psychology of Mineral Wealth:

# Empirical Evidence from Kazakhstan

Elissaios PAPYRAKIS [a]

Osiris Jorge PARCERO [b] [*]

[a] *International Institute of Social Studies (ISS), Erasmus University Rotterdam, Kortenaerkade 12, 2518 AX, The Hague, The Netherlands*

[b] *International School of Economics, Kazakh British Technological University, Almaty, Kazakhstan*

**Declarations of interest:** none

---

[*] Correspondence: Osiris Jorge Parcero: International School of Economics, Kazakh British Technological University, Almaty, Kazakhstan, e-mail: o_parcero@ise.ac; ORCID ID: 0000-0002-6899-7068.

# The Psychology of Mineral Wealth:
# Empirical Evidence from Kazakhstan

## Abstract

Despite rapidly-expanding academic and policy interest in the links between natural resource wealth and development failures – commonly referred to as the 'resource curse' – little attention has been devoted to the psychology behind the phenomenon. Rent-seeking and excessive reliance on mineral revenues can be attributed largely to social psychology. Mineral booms (whether due to the discovery of mineral reserves or to the drastic rise in commodity prices) start as positive income shocks that can subsequently evolve into influential and expectation-changing public and media narratives; these lead consecutively to unrealistic demands that favor immediate consumption of accrued mineral revenues and to postponement of productive investment. To our knowledge, this paper is the first empirical analysis that tests hypotheses regarding the psychological underpinnings of resource mismanagement in mineral-rich states. Our study relies on an extensive personal survey (of 1977 respondents) carried out in Almaty, Kazakhstan, between May and August 2018. We find empirical support for a positive link between exposure to news and inflated expectations regarding mineral availability, as well as evidence that the latter can generate preferences for excessive consumption, and hence, rent-seeking.

## 1. Introduction

There is extensive literature probing the frequency of natural resource mismanagement, especially in the context of mineral-dependent economies (see Papyrakis, 2017 for a review of the literature). A common finding is that mineral-rich states often suffer from low growth rates, excessive macroeconomic volatility, and slow poverty alleviation (in comparison to mineral-scarce countries of similar levels of socio-economic development). The negative impact of mineral wealth is not solely confined to macroeconomic outcomes but also extends to broader development indicators like gender equality, educational and health outcomes, sustainability, life satisfaction, poverty, etc. (For example, see Ali et al., 2020, Apergis and Katsaiti, 2018, Boos and Holm-Müller, 2016, Gearhart and Michieka, 2019, Mignamissi and Malah Kuete, 2021). The phenomenon of underperformance in socio-economic and institutional dimensions, despite abundant mineral rent, is commonly referred to as the resource curse paradox (Colgan, 2014).

In recent years there has been a gradual shift of interest away from standard macroeconomic explanations of natural resource mismanagement (e.g., those based on Dutch Disease theory and debt overhang conditions). Instead, much more attention has been devoted to institutional explanations of the curse. Some of these analyses try to explain how mineral rents allow authoritarian regimes and incompetent leaders to prolong their stays in power (in exchange for patronage, transfers, and favors, see Ross, 2015); similarly, autocrats in mineral-rich regimes are likely to purposively stifle innovation that can potentially remove political power from their hands (Rosenberg and Tarasenko, 2020). More broadly, the presence of abundant mineral rents incentivizes rent-seeking behavior and dissipation of accrued public revenues. Interest groups often vie for larger shares of accrued mineral rents and exert pressure on governments to achieve this (e.g., through strikes, lobbying, or voting behavior, see Baland and Francois, 2000). Especially in ethnically fragmented nations, rent-seeking competition across different groups can result in civil conflict (Elbadawi and Soto, 2015; San-Akca et al., 2020).

Collier (2017) provides a novel psychological explanation for resource mismanagement (and the resource curse) in a recent conceptual paper. He suggests that rent-seeking and excessive consumption of mineral revenues can be attributed mainly to social psychology. Mineral booms (whether due to the discovery of mineral reserves or to drastic increases in commodity prices) start as positive income shocks that subsequently evolve into influential and expectation-changing narratives (for a discussion on the role of narratives in influencing environmental behavior, see Brown, 2017). The dissemination of information on mineral wealth by both media and government agencies may result in unfounded euphoria and unrealistic expectations. Most citizens do not have complete information on what mineral discoveries imply for their personal income (or welfare more broadly). Inflated narratives about the impact of mineral wealth lead to the formation of populist demands that favor immediate consumption (and postponement of investment) of accrued mineral revenues. Determining the origin of demands for excessive consumption is crucial because it may lead to low quality of fiscal performance (and reduced fiscal decentralization), higher debt levels and ultimately lower growth (Ampofo et al., 2021; El Anshasy and Katsaiti, 2013; Wang et al., 2021).

Our analysis aims at empirically testing the psychological foundations of the resource curse, as put forward in a theoretical note by Collier (2017). To our knowledge, this is the first empirical analysis to examine how perceptions of resource rent availability in mineral-rich nations can increase citizen pressure for excessive consumption. We contribute to the literature on the resource curse by examining how perceptions (overestimations) of mineral wealth can be influenced by exposure to associated news, and also by personal interest and characteristics (e.g., level of education, age, etc.). Our study relies on an extensive personal survey (of 1977 respondents) carried out in Almaty, Kazakhstan, between May and August 2018. We focus on Kazakhstan, both due to its extensive extractive sector and its high vulnerability to the resource curse, primarily due to the presence of weak institutions (see Biresselioglu et al., 2019 for their newly-developed resource curse vulnerability index and the corresponding ranking of mineral-rich countries). We indeed find empirical support for Collier's hypotheses with respect to the positive link between exposure to news (and resource-related narratives) and inflated expectations about mineral availability. We also find that the latter can generate preferences for excessive consumption. We expect these findings to be more broadly applicable to other mineral-dependent countries with similar characteristics (i.e., weak institutions, government mistrust, large inequality, etc.).

The structure of the paper is as follows. Section 2 provides a theoretical note on psychological biases created by mineral wealth. Section 3 describes our dataset. Section 4 explains our methodological approach and key findings. Section 5 concludes.

## 2. A theoretical note on the psychology of mineral wealth

Several scholars researching the developmental failures of mineral-rich states emphasize an overall tendency in these nations to prioritize consumption over saving and investing for the future. To a certain extent, this may result from short-sighted government policies designed either to favor local elites and certain interest groups or to broaden public support and prolong periods in power (Frynas and Buur, 2020). Mineral booms also generally create excessive optimism, with governments overestimating their ability to sustain future income and consumption levels based on use of mineral rents (as opposed to investments of the rents, Papyrakis and Gerlagh, 2006). On many occasions, governments have financed reckless spending not only by reliance on mineral income but also through excessive borrowing, using mineral resources as collateral. Such procyclical policies – where governments increase spending and transfers during mineral booms – are often perceived as risky by foreign investors, given extreme volatility in mineral markets and prices. In the long term, such strategies exacerbate overall business cycle fluctuations and heighten the risk of future defaults on international payment obligations (Badeeb et al., 2017).

Collier (2017) presents an innovative theoretical framework to explain possible psychological foundations underlying resource mismanagement and the temptation to overconsume mineral rents. He identifies several potential psychological drivers behind citizens' exaggerated expectations of mineral wealth and associated pressure for excessive consumption. First, exposure to memorable, high-profile narratives (e.g., in the form of powerful messages appearing as newspaper headlines or news broadcasting) can distort perceptions of reality and generate unrealistic expectations (see Jia et al. 2020). Audiences often react to such narratives in a rather passive manner without bringing critical reflection to bear on the validity of information presented. Second, unrealistic expectations (and demands) can also be driven by optimism in forecasting future mineral revenues. Multiple layers of uncertainty characterize mining projects (e.g., concerning future commodity prices, geological features, changes in local and global socio-economic and political conditions, exploration rights, execution timelines, etc.). Both the media and politicians frequently present best-case scenarios as baselines, with little reference to associated complexities and uncertainties (Collier, 2017, and Weszkalnys, 2008). Third, companies active in the extractive sector often exaggerate not only the extent of reserves and possible future revenues, but also the commercial viability of their operations. This strategy enables them to access easier and cheaper external funding, especially in socio-political environments generally characterized as risky. At the same time, such behavior feeds into the unrealistic expectations of citizens regarding the benefits they anticipate from mining activities (see also Gilbert, 2020).

Strong citizen preference for immediate consumption, rather than investment, of mineral rents may be driven by a widespread feeling of distrust toward politicians and government institutions. In many mineral-rich countries, kleptocratic governments have long histories, with the result that the public often fears that self-interested politicians will appropriate mineral revenues for their own benefit. The public reasons that immediate consumption will reduce the risk that mineral rents may disappear into the pockets of officials, preventing benefit from accruing to the average citizen (Fenton Villar, 2020). It is also the case that most mineral-dependent economies tend to be characterized by extensive poverty and lack of government ability to meet basic citizen needs (Collier, 2007). Within such contexts, citizens tend to have relatively high rates of time preference (impatience), favoring immediate consumption over investment and uncertain future benefits (see Adonteng-Kissi, 2017, and Lawrance, 1991).

To reverse the psychological drivers of unrealistic expectations and demands for consumption, Collier (2017) puts forward several recommendations. Given the optimism that mineral wealth produces when presented in aggregate numbers (and the inability of many to grasp the implications of these coarse figures), presenting any projected monetary flows (e.g., in terms of expected revenues) in per capita terms would reduce overall psychological bias (see Collier, 2017, as well as Cust et al., 2017). Equally important is the need for government officials to explain the rationale behind favoring the allocation of mineral rents to investment rather than immediate consumption. Citizens will accept a 'sacrifice' if they become convinced that investment of mineral revenues (e.g., in public infrastructure, or health/educational projects) is a strategy with larger long-term (and sustained) benefits (Papyrakis and Gerlagh, 2006, Edmunson, 2014). Naturally, all this also requires that politicians and government officials try to build citizen trust in them. By itself, such an exercise requires painstaking, long-term processes, during which government officials need to adopt a consistent political ethos based on modesty and transparency (and hence, to lead by example).

## 3. Survey description and data on perceived mineral wealth

For our study, we carried out an extensive survey of 1977 respondents in the center of Almaty, Kazakhstan, between May and August 2018. The survey was conducted in Russian and Kazakh (the official languages of Kazakhstan). Appendix 1 provides a summary of the questionnaire (in English).

The focal question of the survey was one referring to each respondent's perceptions about the country's mineral wealth. For Kazakhstan, this meant three primary mineral resources: gas, oil, and coal.

For all three resources, respondents were asked to reflect on the value of Kazakhstan's reserves. During our pilot study preceding the survey, it became evident that most respondents had real difficulties in coming up with estimates regarding the value of subsoil mineral assets. As a result, it was necessary to provide guidance to survey participants to get meaningful answers, while at the same time taking precautions to minimize any 'anchoring effect' that would bias our results (i.e., a cognitive bias where a participant's response would depend too heavily on information offered in the survey). For example, in the case of gas reserves, each respondent was provided with a ranking of 206 countries, ranging from the most gas-abundant nations to the most gas-scarce ones. The monetized value of each country's gas reserves for 2017 (in trillions of tenge, Kazakhstan's local currency) was provided next to each country's name. Kazakhstan was purposely excluded from the list. Data on international gas reserves came from the U.S. Energy Information Administration (2018); for our calculations, we used the average 2017 international price for gas ($0.0028 per cubic feet, IEA, 2018). Each respondent was then requested to reflect on where Kazakhstan should be located in the international ranking of nations with gas reserves, and then estimate Kazakhstan's gas wealth (hereafter referred to as 'surveyee's perception of gas wealth'). For the sake of demonstration, Appendix 2 presents part of the international rankings for gas reserves (for the most gas abundant nations). In effect, Kazakhstan's gas reserves were worth about 76 trillion tenge at the time of the survey (and were, hence, located between Mozambique and Egypt in terms of relative importance – with the 15$^{th}$ largest reserves globally).

Similarly, respondents were provided with country rankings for oil and coal reserves and were asked to reflect on Kazakhstan's position in relation to other economies and the corresponding value of its oil/coal assets. Data on international oil and coal reserves again came from the U.S. Energy Information Administration (2018); for our calculations, we made use of the average 2017 international price for oil and coal ($66 per barrel of oil and $70 per ton of coal, IEA, 2018). At the time of the survey, Kazakhstan's had the 12th largest oil reserves globally (70 billion barrels, equivalent to approximately 634 trillion tenge) and the 8th largest coal reserves (25.6 billion tons, equivalent to about 574 trillion tenge).

## 4. Results

The survey data allows us to measure the extent to which respondents in Kazakhstan overestimated the actual availability of the country's mineral wealth. As a first step in our analysis, we constructed an index of this overestimation per type of mineral reserves *i* (i.e., gas, oil, coal) for each survey participant *j*:

$$\text{Surveyee's overestimation rate of the country's mineral wealth}_{i,j} = \frac{\text{Surveyee's perception of mineral wealth}_{i,j} - \text{Real mineral wealth}_i}{\text{Real mineral wealth}_i} \quad (1)$$

Any value above 0 corresponds to an overestimation of the matching mineral asset. We found strong evidence of an overestimation of the country's mineral wealth based on responses of the 1977 survey participants. The average overestimation rate for Kazakhstan's gas, oil, and coal wealth was equal to 4.99, 1.99, and 1.23 respectively. In other words, on average, respondents believed that Kazakhstan had gas, oil, and coal reserves that were worth 499%, 199%, and 123% more than their actual values! Descriptive statistics (for all dependent and explanatory variables) are presented in Appendix 3; Appendix 4 provides a correlation table.

Next, we examined how exposure to news, which in Kazakhstan typically publicizes important, high-profile developments in the local extractive sector, could generate unrealistic expectations of mineral wealth. We proposed the following hypothesis:

**Hypothesis 1**: Greater exposure to news items is associated with larger overestimation rates of Kazakhstan's mineral wealth.

Table 1 applies OLS regressions to test the hypothesis. The dependent variable is the surveyee's overestimation of the country's mineral wealth (for gas, oil, and coal, respectively) as described above. In all the regressions, we include the natural logarithm of a respondent's interest in news items as an explanatory variable (*Ln_int_news*). The first three parsimonious regressions include two additional regressors, namely the respondent's self-assessment of optimism (*Ln_optimism*) and his/her extent of interest in economic issues (*Ln_int_econ*). We anticipated a positive correlation between expressed optimism and the extent of a respondent's overestimation of the country's mineral wealth (given the general disposition of optimists to look on the more favorable side of events). On the other hand, we expected a negative correlation between interest in economic issues and overestimation of the extent of mineral wealth (given that those better informed on economic matters are likely to be less prone to biased overestimation based on influential reports and news items. There is considerable research in experimental and behavioral economics linking biases, including those concerning economic/quantitative assessments, and interest in economic issues (see Slonim et al., 2013, who claim that interest in economics increases reflection and effort dedicated to understanding the purposes and design of experimental economic lab activities, and also Wright, 2010, who links interest in economics with an enhanced understanding of pension schemes). All three explanatory variables (measuring the extent of optimism, interest in news items, and interest in economic issues) are measured initially on a scale from 1 (very low) to 5 (very high).

Our key focus is on the relationship between a respondent's exposure to news items (as captured by *Ln_int_news*) and his/her overestimation of Kazakhstan's mineral wealth. Results accord with intuition and support Collier's hypothesis: as can be seen in regression 1, increased exposure to news correlates positively with a higher overestimation of the country's gas wealth (statistically significant at the 1% level). Very similar results are obtained for oil (regression 2). In the case of coal, there is a positive but statistically insignificant correlation (regression 3). The magnitude of the effect is quite prominent in the case of gas and oil, as measured by corresponding elasticities presented at the bottom of Table 1. A 1% rise in the index measuring the extent of news exposure is associated with a 0.44% (0.47%) increase in the overestimation of the country's gas (oil) wealth. The smaller (and statistically insignificant) elasticity for coal may be attributed to the following two factors. First, coal is mainly consumed domestically for electricity production rather than used for exports (unlike gas and oil); consequently, coal receives less media attention. Second, while Kazakhstan has vast coal reserves (almost on par with its oil reserves when expressed in monetary terms), the contribution of coal rents to annual GDP values is relatively modest (0.9% for 2017, against 10.2% and 1.2% for oil and gas respectively, see World Development indicators, 2020).

Regressions (1)-(3) also reveal that, as posited earlier, the more optimistic a person is, the higher his/her overestimation of the country's mineral wealth will be. The coefficient of interest in economics is also in accordance with our earlier expectation: those better informed on economic issues are less likely to overestimate the extent of gas/oil/coal wealth.

Table 1. Determinants of overestimation rates per type of natural wealth (gas, oil, and coal)

|  | (1) Gas | (2) Oil | (3) Coal | (4) Gas | (5) Oil | (6) Coal | (7) Gas | (8) Oil | (9) Coal |
|---|---|---|---|---|---|---|---|---|---|
| *Ln_int_news* | 2.17*** | 0.93*** | 0.28 | 1.96*** | 0.85*** | 0.26 | 1.77*** | 0.81*** | 0.32* |
|  | (0.61) | (0.18) | (0.19) | (0.62) | (0.18) | (0.19) | (0.65) | (0.19) | (0.19) |
| *Ln_optimism* | 2.39*** | 1.01*** | 0.85*** | 2.01*** | 0.92*** | 0.80*** | 2.08*** | 0.93*** | 0.83*** |
|  | (0.58) | (0.20) | (0.17) | (0.57) | (0.20) | (0.17) | (0.61) | (0.21) | (0.17) |
| *Ln_int_econ* | -2.03*** | -0.61*** | -0.39*** | -2.59*** | -0.86*** | -0.40** | -2.73*** | -0.93*** | -0.37** |
|  | (0.40) | (0.14) | (0.13) | (0.48) | (0.18) | (0.16) | (0.48) | (0.18) | (0.17) |
| *Ln_int_pol* |  |  |  | -0.04 | 0.16 | -0.12 | 0.01 | 0.12 | -0.06 |
|  |  |  |  | (0.51) | (0.17) | (0.15) | (0.52) | (0.17) | (0.15) |
| *Ln_int_buss* |  |  |  | 2.35*** | 0.61*** | 0.34** | 2.50*** | 0.64*** | 0.22 |
|  |  |  |  | (0.46) | (0.19) | (0.15) | (0.48) | (0.19) | (0.15) |
| *Ln_corruption* |  |  |  |  |  |  | 0.49 | 0.27 | 0.19 |
|  |  |  |  |  |  |  | (0.49) | (0.21) | (0.19) |
| *Work* |  |  |  |  |  |  | -0.37 | 0.19* | 0.01 |
|  |  |  |  |  |  |  | (0.31) | (0.11) | (0.10) |
| *Ln_age* |  |  |  |  |  |  | -0.05 | 0.22 | -0.26* |
|  |  |  |  |  |  |  | (0.50) | (0.17) | (0.14) |
| *Ln_education* |  |  |  |  |  |  | 1.96** | 0.43 | 0.08 |
|  |  |  |  |  |  |  | (0.95) | (0.36) | (0.35) |
| *Constant* | 1.19 | 0.08 | 0.15 | -0.42 | -0.37 | -0.06 | -4.01* | -2.24*** | 0.37 |
|  | (0.75) | (0.31) | (0.28) | (0.78) | (0.33) | (0.30) | (2.21) | (0.72) | (0.75) |
| *Observations* | 1,965 | 1,966 | 1,966 | 1,965 | 1,966 | 1,966 | 1,898 | 1,899 | 1,899 |
| *R-squared* | 0.04 | 0.05 | 0.02 | 0.06 | 0.06 | 0.03 | 0.06 | 0.06 | 0.03 |
| Elasticities for key statistically significant variables | | | | | | | | | |
| *Int_news* | 0.44*** | 0.47*** | 0.22 | 0.39*** | 0.43*** | 0.21 | 0.36*** | 0.41*** | 0.26* |
| *Optimism* | 0.48*** | 0.51*** | 0.69*** | 0.40*** | 0.46*** | 0.65*** | 0.42*** | 0.47*** | 0.68*** |
| *Int_econ* | -0.41*** | -0.31*** | -0.32*** | -0.52*** | -0.43*** | -0.33** | -0.55*** | -0.47*** | -0.30** |
| *Int_buss* |  |  |  | 0.12*** | 0.08*** | 0.07** | 0.13*** | 0.08*** | 0.05* |

Note: Robust standard errors in brackets. Superscripts *, **, and *** correspond to a 10, 5, and 1% significance level. The elasticities are measured at mean values.

It might be the case that the negative association between overestimating a country's natural wealth and interest in economics is not exclusive to this disciplinary field. If the influence were not exclusive, this would weaken our previous argument that attributes a smaller bias in estimations solely to interest in economics. For this reason, we evaluate whether interests in politics and business might also be linked to the overestimation rate, since both are related to public affairs. Economics and business are, however, different in a fundamental respect: evidence suggests that people interested in business are typically more optimistic than the average person (Cooper et al., 1988; Ucbasaran et al., 2010). Thus, among other things, interest in business issues may capture that part of optimism that is not directly controlled by the variable *Ln_optimism* (self-assessed optimism), at least to the extent that optimistic people may not be fully aware of their optimism.

We add the variables *Ln_int_pol* and *Ln_int_buss* to capture the extent of expressed interest in political and business issues, respectively (again measured on a scale from 1 - very low - to 5 - very high). Evidence that interest in business issues indicates optimism can be seen in the positive correlation between the variables *Ln_int_buss* and *Ln_optimism*, 0.21 (see Appendix 4). This is the highest pairwise correlation of *Ln_int_buss* with any other variable, about twice the correlation coefficient between *Ln_int_econ* and *Ln_optimism* (0.11).

The results in columns (4)-(6) of Table 1 show that the relationship between one's exposure to news items (*Ln_int_news*) and his/her overestimation of Kazakhstan's mineral wealth remains robust to the addition of the variables *Ln_int_pol* and *Ln_int_buss*. The corresponding coefficients for the variable *Ln_int_news* decrease only slightly and remain statistically significant in the case of gas and oil. The measures of elasticity at the bottom of Table 1 show that a 1% rise in the index measuring the extent of news exposure is associated with a 0.39% (0.43%) increase in overestimation of the country's gas (oil) wealth. Interest in politics consistently appears statistically insignificant, and as expected, the variable *Ln_int_buss* shows a positive and significant effect in the three specifications.

In columns (7)-(9), we include four additional control variables potentially influencing perceptions of Kazakhstan's mineral wealth. We include the perception of corruption regarding management of public resources (*Ln_corruption*: scale 1 to 5) and a dummy variable taking the value of 1 if a respondent is currently employed (*Work*). In addition, we include the natural logarithm of the respondent's age (*Ln_age*) and level of education (*Ln_education*: scale 1 to 8, see Appendix 1 for the corresponding levels of educational attainment). The extent of perceived corruption is expected to be positively correlated with one's overestimation of a country's mineral wealth. Respondents who acknowledge corruption as a serious concern in Kazakhstan are likely to believe that part of the country's mineral wealth is embezzled, and hence, that it is intentionally under-reported (i.e., actual mineral wealth is higher than that reported by official sources). Regressions (7)-(9) indeed present a positive coefficient for corruption, albeit a non-statistically significant one. The variables *Work*, *Ln_age,* and *Ln_education* are only significant for oil, coal, and gas regressions, respectively. Most importantly, the earlier results regarding a positive relationship between one's exposure to news items (*Ln_int_news*) and his/her overestimation of Kazakhstan's mineral wealth still hold (with the corresponding coefficient for coal now also becoming statistically significant, albeit at the 10% level).

Our second key hypothesis focuses on how inflated expectations may influence preferences regarding how mineral rents should be spent (see also Collier, 2017):

**Hypothesis 2**: An overestimation of the country's mineral wealth is likely to generate unrealistic expectations and pressure for excessive redistribution and consumption.

We asked respondents to a.) provide an estimated guess on the current division of mineral revenues between immediate consumption ('spend money for people's satisfaction today', e.g., in the form of subsidies, wages of public servants, income transfers) and saving/investment (e.g., in the form of infrastructure development, training programs, purchase of productive equipment, etc.) and b.) reveal their preferences regarding their ideal division of revenues between these two uses. For the purpose of proxying demand for excessive consumption, we constructed a new index based on the ratio of ideal versus actual allocation:

$$\ln\left(\genfrac{}{}{0pt}{}{Pressure\ for}{excessive\ consumption}\right) = \ln\left(\frac{Govt.should\ spend\ (consumption/investment)}{Govt.spends\ (consumption/investment)}\right). \qquad (2)$$

For those respondents who felt that the government should allocate more (fewer) resources toward consumption, the ratio becomes greater (less) than 1, and the corresponding natural logarithm larger (smaller) than 0. The average (logarithmic) value of the index is 0.45, which means that the average preferred ratio of consumption/investment vs. the actual one is close to 1.57. We test *hypothesis 2* through a series of regressions presented in Table 2.

Table 2. Determinants of citizen pressure for excessive consumption

| VARIABLES | (1) | (2) | (3) |
|---|---|---|---|
| Overestimation_total | 0.03*** | 0.03*** | 0.03*** |
|  | (0.01) | (0.01) | (0.01) |
| Ln_Corruption | 0.11* | 0.07 | 0.07 |
|  | (0.05) | (0.06) | (0.06) |
| Ln_Optimism |  | 0.05 | 0.05 |
|  |  | (0.06) | (0.06) |
| Ln_age |  | 0.20*** | 0.21*** |
|  |  | (0.04) | (0.05) |
| Work |  |  | -0.04 |
|  |  |  | (0.04) |
| Ln_Education |  |  | -0.03 |
|  |  |  | (0.12) |
| Constant | 0.08 | -0.64*** | -0.61*** |
|  | (0.08) | (0.19) | (0.22) |
| Observations | 1,398 | 1,387 | 1,387 |
| R-squared | 0.01 | 0.02 | 0.03 |

Note: Robust standard errors in brackets. Superscripts *, **, and *** correspond to a 10, 5, and 1% significance level.

The critical focus of Table 2 (to test *hypothesis 2*) lies in the relationship between a respondent's degree of pressure for excessive consumption and his/her overestimation of mineral wealth. In all regressions, the primary variable is *Overestimation_total*, which measures the aggregate overestimation of wealth for gas, oil, and coal. It is calculated using expression (1), where subscript $i$ now indicates the aggregate value of gas, oil, and coal. Regression 1 shows that there is indeed a positive and statistically significant (at the 1% level) correlation between the two variables. Expressed in elasticity terms, an increase of the overestimation rate by 1% corresponds to an increase in pressure for excessive consumption (the ratio inside the bracket in expression (2)) of 0.06%, when the overestimation rate is measured at its mean value. In the same specification, we have also added the perception of corruption regarding mismanagement of public resources (*Ln_Corruption*). Collier (2017) suggests that other things equal, mistrust in government management makes citizens prefer more immediate consumption of public resources instead of the more uncertain benefits accruing from investment, which may never materialize. The coefficient for the variable perception of corruption is positive and statistically significant at the 10% level (although it loses its statistical significance for the richer specifications that follow). Thus, an increase in perception of corruption by 1% corresponds to an increase in pressure for excessive consumption by 0.105%.

In column (2) of Table 2, we add two more control variables, namely the respondent's extent of optimism and age (*Ln_optimism* and *Ln_age*). One may suspect that optimistic respondents are more inclined to envisage brighter prospects for the future (e.g., see Dawson and Henley, 2012), and hence, to see less need for investment at the expense of immediate consumption. The coefficient is indeed positive but insignificant. On the other hand, age has a positive and statistically significant coefficient; this may suggest that the elderly (being closer to the ends of their lives) have a stronger preference for consumption than for investment and especially so in the case of public resources, which they cannot directly bequeath to their offspring (see also Modigliani, 1966, and Hurd, 1990). The relationship between the overestimation rate and the index of pressure for excessive consumption (the focal point of Table 2) remains positive, of similar size, and statistically significant. In column (3), we enrich our specification with two additional control variables: the dummy variable capturing employment (*Work*) and the variable measuring level of education (*Ln_education*). One may surmise that employed individuals favor saving/investment as they face fewer budget constraints. Indeed, the corresponding coefficient for *Work* is negative but statistically insignificant. Educated individuals may also face fewer budget constraints, as they are likely to have higher income levels and hence favor saving/investment. While the sign of the coefficient accords with this intuition, it is statistically insignificant (note that the coefficients for these two control variables remain statistically insignificant even when they are added separately instead of jointly). More importantly, our main results regarding the coefficient of the overestimation index remain robust.

## 5. Conclusion and Policy Implications

Despite keen academic and policy interest in the links between natural resource wealth and development failures, little attention has been devoted to the psychological underpinnings of the resource curse. Mineral booms are typically associated with rent-seeking behavior. Individuals strive for a share of accrued mineral rents and exert pressure on the government to achieve this. In this paper, we used data from extensive fieldwork carried out in Almaty, Kazakhstan, to examine possible psychological foundations underlying mineral rent-seeking. Positive mineral shocks often translate into influential and expectation-changing narratives. We find empirical support for a positive link between exposure to news and inflated expectations of mineral availability. We also find that the latter can generate preferences for excessive consumption (and hence, rent-seeking).

Our research has important policy implications. First, it demonstrates that information clarity is crucial regarding the availability of mineral rents and their potential to transform national economies and individual livelihoods. Governments in mineral-rich countries need to provide accessible information about the relative importance of the extractive sector in ways that are easily comprehensible to all citizens (see also Collier, 2017). Large aggregate numbers (e.g., of the value of reserves discovered) are unlikely to resonate with most citizens and may generate confusion and unrealistic expectations. Expressing values in per capita terms can help prevent unjustified mineral-induced euphoria that is likely to translate into demands for excessive short-term consumption and widespread rent-seeking behavior.

Naturally, the present research is only a first step toward empirically examining the psychological foundations of the resource curse and natural resource mismanagement more broadly. This could be especially interesting in developing countries experiencing sudden major discoveries of mineral resources. In such cases, the research would further benefit if time-series data allows monitoring changes in public opinion due to the discovery.

# Appendices

Appendix 1. Summary of questionnaire

| Variable | Question asked | Possible answers |
|---|---|---|
| Surveyee's perceptions of gas, oil, and coal wealth | What do you think the value of Kazakhstan's gas, oil, and coal wealth is? | Value of gas ______, oil ______, and coal reserves ______ respectively (in tenge, Kazakhstan's national currency) |
| Government allocation (consumption / investment) | In which proportion do you think that the government allocates gas/oil/coal revenues between current consumption ('spend money for people's satisfaction today', e.g. in the form of subsidies, wages of public servants, income transfers) and saving/investment (e.g., in the form of infrastructure development, training programs, purchase of productive equipment etc.)? | Proportion of gas/coal/oil revenues allocated towards<br>Consumption: ___%<br>Saving/Investment: ___%* |
| Ideal government allocation (consumption / investment) | In which proportion do you think the government should allocate gas/oil/coal revenues between current consumption and saving/investment? | Proportion of gas/coal/oil revenues that should be allocated towards<br>Consumption: ___%<br>Saving/Investment: ___%* |
| Int_news | From 1 to 5, how much do you read, listen to or watch the news? | 1  2  3  4  5  (1 not at all, 5 very much) |
| Int_econ | From 1 to 5, how would you rate your interest in economics? | 1  2  3  4  5  (1 very low, 5 very high) |
| Int_pol | From 1 to 5, how would you rate your interest in politics? | 1  2  3  4  5  (1 very low, 5 very high) |
| Int_buss | From 1 to 5, how would you rate your interest in business? | 1  2  3  4  5  (1 very low, 5 very high) |
| Optimism | Do you consider yourself a pessimistic or optimistic person? | 1  2  3  4  5  (1 highly pessimistic, 5 highly optimistic) |
| Corruption | A proportion of public resources becomes lost (wasted) because of inefficiency/corruption at different government levels. How substantial do you think this loss of resources is? | 1  2  3  4  5  (1 very low, 5 very high) |
| Work | Do you work? | Yes/No |
| Age | How old are you? (age in years) | Any number |
| Education | What is the highest level of education that you have completed? | Choose the letter that applies to your case<br>A. PRIMARY SCHOOL INCOMPLETE<br>B. PRIMARY SCHOOL COMPLETED<br>C. SECONDARY SCHOOL INCOMPLETE<br>D. SECONDARY SCHOOL COMPLETED<br>E. UNIVERSITY INCOMPLETE<br>F. UNIVERSITY COMPLETED<br>H. POSTGRADUATE DEGREE INCOMPLETE<br>G. POSTGRADUATE DEGREE COMPLETED |

NOTE: * The reason for asking for both shares was to ensure that respondents understood that the two percentages need to add up to 100%. This was added after a pilot survey showed that asking for only one share generated confusion for many respondents.

Appendix 2. International ranking of gas reserves (value in trillion tenge) for the 20 most gas abundant countries (excluding Kazakhstan)

|    | Country | trillion tenge |
|----|---------|---------------:|
| 1  | Russia | 1,513 |
| 2  | Iran | 1,067 |
| 3  | Qatar | 762 |
| 4  | United States | 289 |
| 5  | Saudi Arabia | 273 |
| 6  | Turkmenistan | 237 |
| 7  | United Arab Emirates | 193 |
| 8  | Venezuela | 182 |
| 9  | Nigeria | 173 |
| 10 | China | 172 |
| 11 | Algeria | 143 |
| 12 | Iraq | 121 |
| 13 | Indonesia | 91 |
| 14 | Mozambique | 90 |
| 15 | Egypt | 69 |
| 16 | Canada | 65 |
| 17 | Australia | 63 |
| 18 | Uzbekistan | 58 |
| 19 | Kuwait | 56 |
| 20 | Norway | 56 |

NOTE: The full list of countries was included in the survey but not reported here.

Appendix 3. Descriptive statistics

| Variable | Obs. | Mean | Std. Dev. | Min | Max |
|---|---|---|---|---|---|
| *Surveyee's overestimation rate of the country's gas wealth* | 1,974 | 4.99 | 6.27 | -1.00 | 72.97 |
| *Surveyee's overestimation rate of the country's oil wealth* | 1,975 | 1.99 | 2.34 | -0.98 | 38.46 |
| *Surveyee's overestimation rate of the country's coal wealth* | 1,975 | 1.23 | 2.05 | -0.99 | 11.20 |
| *ln(pressure for excessive consumption)* | 1,950 | 0.45 | 0.84 | -2.20 | 4.60 |
| *Overestimation_total* | 1,973 | 1.82 | 1.94 | -0.97 | 20.02 |
| *Ln_int_news* | 1,972 | 1.35 | 0.32 | 0.00 | 1.61 |
| *Ln_optimism* | 1,972 | 1.37 | 0.30 | 0.00 | 1.61 |
| *Ln_int_econ* | 1,973 | 1.18 | 0.41 | 0.00 | 1.61 |
| *Ln_int_pol* | 1,973 | 1.14 | 0.43 | 0.00 | 1.61 |
| *Ln_int_buss* | 1,974 | 1.32 | 0.35 | 0.00 | 1.61 |
| *Ln_corruption* | 1,924 | 1.41 | 0.29 | 0.00 | 1.61 |
| *Ln_age* | 1,977 | 3.57 | 0.37 | 0.00 | 1.00 |
| *Work* | 1,961 | 0.66 | 0.47 | 2.56 | 4.41 |
| *Ln_education* | 1,977 | 1.76 | 0.17 | 0.00 | 2.08 |

Appendix 4. Correlation table

|  |  | (1) | (2) | (3) | (4) | (5) | (6) | (7) | (8) | (9) | (10) | (11) | (12) | (13) | (14) |
|---|---|---|---|---|---|---|---|---|---|---|---|---|---|---|---|
| Surveyee's overestimation rate (gas) | (1) | 1.00 | | | | | | | | | | | | | |
| Surveyee's overestimation rate (oil) | (2) | 0.40 | 1.00 | | | | | | | | | | | | |
| Surveyee's overestimation rate (coal) | (3) | 0.36 | 0.43 | 1.00 | | | | | | | | | | | |
| ln(pressure for excessive consumption) | (4) | 0.16 | 0.17 | 0.10 | 1.00 | | | | | | | | | | |
| Overestimation_total | (5) | 0.59 | 0.87 | 0.79 | 0.18 | 1.00 | | | | | | | | | |
| Ln_int_news | (6) | 0.14 | 0.18 | 0.09 | 0.05 | 0.17 | 1.00 | | | | | | | | |
| Ln_optimism | (7) | 0.14 | 0.16 | 0.14 | 0.04 | 0.19 | 0.33 | 1.00 | | | | | | | |
| Ln_int_econ | (8) | -0.11 | -0.07 | -0.04 | -0.05 | -0.08 | 0.16 | 0.11 | 1.00 | | | | | | |
| Ln_int_pol | (9) | -0.04 | 0.01 | -0.02 | 0.01 | -0.01 | 0.20 | 0.09 | 0.60 | 1.00 | | | | | |
| Ln_int_buss | (10) | 0.12 | 0.11 | 0.05 | 0.10 | 0.11 | 0.19 | 0.21 | 0.32 | 0.24 | 1.00 | | | | |
| Ln_corruption | (11) | 0.01 | 0.03 | 0.02 | 0.04 | 0.03 | 0.04 | -0.05 | 0.02 | 0.05 | -0.02 | 1.00 | | | |
| Ln_age | (12) | -0.02 | 0.05 | 0.00 | -0.02 | 0.03 | 0.03 | 0.00 | 0.10 | 0.04 | 0.13 | 0.03 | 1.00 | | |
| Work | (13) | 0.02 | 0.07 | -0.04 | 0.11 | 0.02 | 0.20 | -0.01 | 0.08 | 0.14 | -0.07 | 0.12 | 0.09 | 1.00 | |
| Ln_education | (14) | 0.04 | 0.06 | -0.02 | 0.06 | 0.04 | 0.13 | 0.01 | 0.10 | 0.11 | 0.05 | 0.09 | 0.25 | 0.45 | 1.00 |